# Lunar Eclipse of June, 15, 2011: Three-color umbra surface photometry

Oleg S. Ugolnikov[1], Igor A. Maslov[1,2], Stanislav A. Korotkiy[3]

[1]Space Research Institute, Russian Academy of Sciences, Russia
[2]Sternberg Astronomical Institute, Moscow State University, Russia
[3]Scientific Center "Ka-Dar", Russia

Corresponding author e-mail: ougolnikov@gmail.com

The paper contains the result description of the photometric observations of the Moon surface during the total lunar eclipse of June, 15, 2011 conducted in southern Russia and Ukraine in three narrow spectral bands with effective wavelengths equal to 503, 677, and 867 nm. The photometric maps of umbra are built, the radial dependencies of relative brightness are compared with the theoretical ones for gaseous atmosphere. Dark anomalies of umbra are shown and related with the aerosol air pollution of the definite regions above the surface of Earth.

## 1. Introduction

Lunar eclipses are the object of theoretical and observational interest over the centuries and until the present time [1, 2]. The radiation coming to the Moon during the totality is transferred through the dense layers of the atmosphere of the Earth. The atmospheric path length exceeds thousand kilometers, and the emission is sufficiently refracted, scattered, and absorbed. All processes strongly depend on the atmospheric conditions, presence of aerosol [3, 4] and optically active gases ($O_3$, $NO_2$, $H_2O$ [5], etc., if observations are held in the absorption spectral bands of these gases). This makes the lunar eclipse observations the tool for the atmospheric optical components investigations.

Theoretical analysis [6] had shown that the scattered fraction of emission reaching the Moon's surface is noticeable only in the case of strong volcanic atmosphere perturbation (as after the Mt. Pinatubo eruption in 1992) or at the wavelengths shorter than 400-450 nm. In the other cases the observed emission is the direct solar radiation refracted in the atmosphere. The lower the path of light propagation above the surface of Earth, the stronger the refraction and the deeper the umbra region where this emission can be detected. Holding the surface photometry of the Moon, we cover the interval of altitudes in troposphere and lower stratosphere and wide range of locations along the Earth's limb where refraction occurs. Variations of atmosphere conditions along the limb can lead to the radial asymmetry of the umbra with the dark spots far from the center, especially in infrared part of spectrum, as it was observed during the eclipse of March, 4, 2007 and related with equatorial region with increased aerosol and water vapor concentration above Indonesia [5].

The basic light extinction mechanism in the atmosphere is the Rayleigh scattering that strongly depends on the wavelength ($\sim\lambda^{-4}$). Short-wave emission disappears traveling along the low trajectory above the Earth's surface, but the longer-wave radiation can pass the atmosphere by almost the same trajectory and reach the Moon. It is the explanation of red color of the eclipsed Moon and choice of red and infrared spectral bands for most of the eclipse photometric observations. If the observations are hold in a number of different spectral bands, it helps to separate the definite absorption component (for example, the water vapor [5]) or to build the wavelength dependency of atmospheric aerosol extinction above the limb, that will be done in this work.

## 2. Observations

Lunar eclipse of June, 15, 2011 was the deepest total eclipse of the first two dozens of years of 21$^{st}$ century, having the maximum magnitude 1.71. The Moon was crossing the umbra close to its diameter almost horizontally from the west to the east. Optical structure of the eclipse was basically formed by equatorial and tropical atmosphere above the equatorial Atlantic ocean and north-west of Africa (in the beginning of totality) and China & western Pasific ocean (in the end of totality).

The photometric measurements of the eclipsed Moon were held close to local midnight in Crimean Laboratory of Sternberg Astronomical Institute (Crimea, Ukraine, 44.7°N, 34.0°E) and Special Astrophysical Observatory (SAO, Russia, 43.7°N, 41.4°E). The observations in Crimea were conducted by CCD-camera with narrow spectral IR-filter with effective wavelength equal to 867 nm. SAO observations were conducted by two CCD-cameras with narrow filters with the effective wavelengths 503 and 677 nm. Both SAO spectral ranges fall on the different slopes of Chappuis bands of ozone absorption, and Crimean IR-range is out of atmospheric gases absorption bands.

Unfortunately, weather conditions restricted the totality observation periods and umbral area covered by measurements. Atmospheric transparency was stable and suitable for observations during the initial 15 minutes of totality in all spectral bands. In addition, Crimean observations at 867 nm had also covered last 35 minutes of totality. Owing to large angular size of the Moon, it allowed to hold the IR-measurements on the majority of umbral area passed by the Moon. The observational data were calibrated by the measurements of nearby standard stars and lunar surface observed right before or after the eclipse. Data obtained at 677 and 867 nm were also used to correct the effects of weak “red tail” of 503 nm filter.

## 3. Results and Discussion

The basic measured value is the brightness ratio of the lunar surface element in the definitive point of the umbra and the same surface element outside the umbra and penumbra. The procedure of ratios calculation with account of variable sky background from brighter parts of the Moon is described in [4]. The distribution maps of measured value for three spectral bands are shown in Figure 1. As it was told above, measurements had covered the majority of umbra for 867 nm and one western area for both 503 and 677 nm bands.

The most remarkable feature seen in the 867 nm map is the sufficient asymmetry of the umbra. While western parts show the regular structure and typical relative brightness values observable during the other eclipses in this band [5, 7], eastern part is dark as a whole, especially to the North-East from the center. The solar radiation transfers to this place through the middle troposphere (about 5-6 km and above) of the eastern China, and the spot can be related with the aerosol pollution of this atmosphere region. The brightness difference of eastern and western umbra parts was noticed by many observers worldwide [8], and the brightness values recorded during the second half of totality at 867 nm were the lowest among the observed eclipses of the last years [7]. It does not completely follow the assumption of French astronomer A. Danjon in 1920s, who had stated the gradual increase of lunar eclipses brightness during and after the solar activity maximum epoch.

At the shorter waves, 677 and especially 503 nm, the western umbra brightness is smaller. It is native since lower atmosphere layers have less influence on the umbra brightness due to their lower transparency. The emission background is basically formed by the outer solar edge radiation transferring above. The 503 nm relative brightness value falls down to about $10^{-6}$ in the umbra part covered by observations, but it is still higher than the theoretically estimated scattering fraction contribution (about $10^{-7}$, [6]).

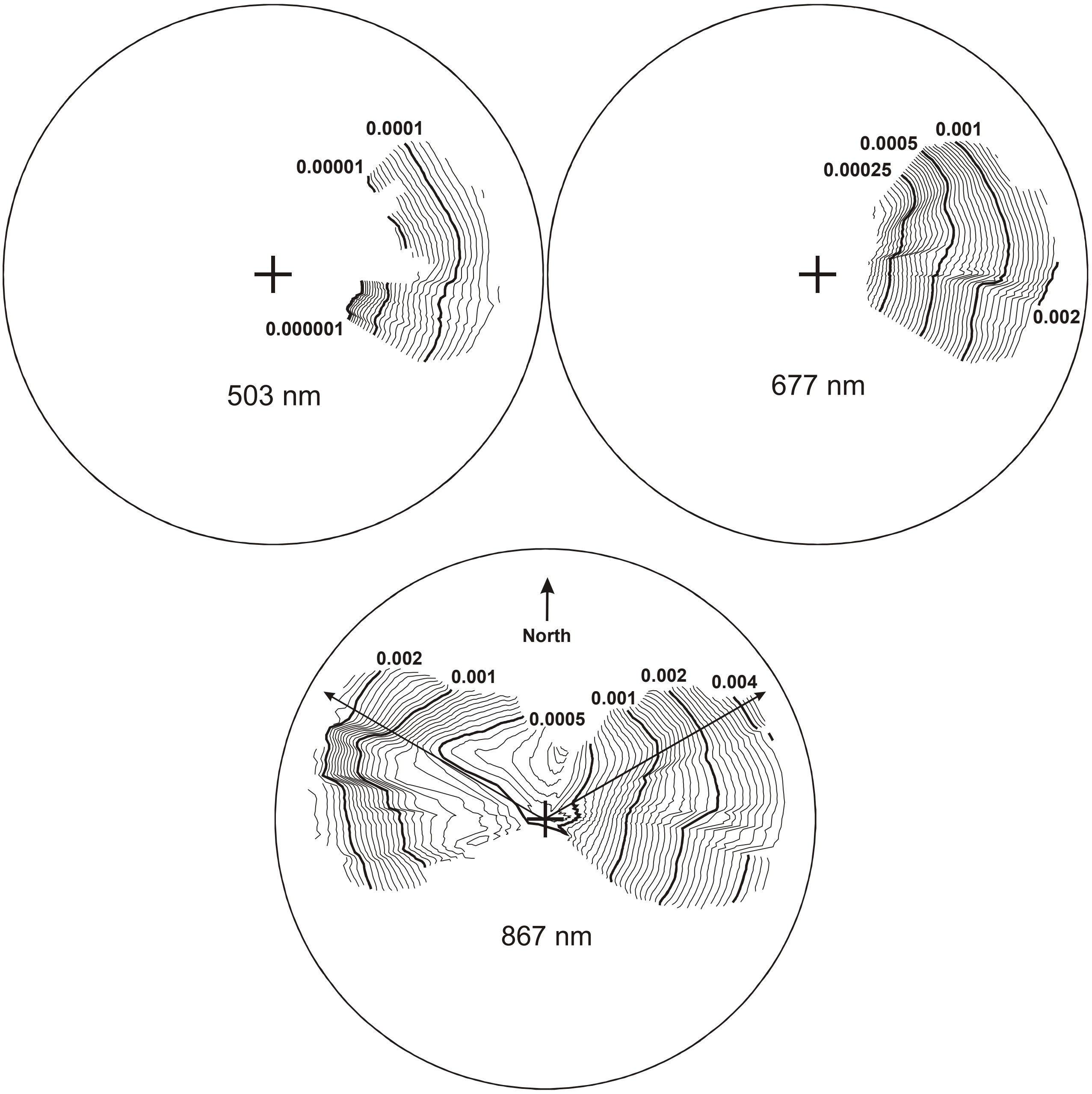

*Figure 1. Relative brightness of Moon surface inside the umbra in three spectral bands. The step between the neighbor isophotes is $10^{0.1}$ for 503 nm and $2^{0.1}$ for other bands.*

Figure 2 shows the radial dependencies of umbra relative brightness for the directions shown by arrows in the Figure 1 (30° to the north from equator). They are signed as W and E for 867 nm, and only western ones are available for other wavelengths. The dependencies are compared with the ones calculated using simple equatorial gaseous atmosphere model (ground temperature 300 K, lower stratosphere temperature 220 K, and total ozone content 250 Dobson units).

Observational curves for outer western umbra (0.6° from the center) are close to the theoretical ones in all three bands, showing the clear conditions of Atlantic upper troposphere. Aerosol extinction appears below, causing the difference between observational and theoretical curves. On the contrary, western umbra is darker from the center to the edge, that can be related with the increased tropospheric aerosol concentration along the eastern limb over the wide range of altitudes. Umbra becomes brighter just at the south edge of the lunar path. This umbra region was emitted by the solar radiation transferred above the Pasific ocean far from Asian shores. The maritime atmosphere (as an Atlantic one for the western umbra half) contains less amount of aerosol particles.

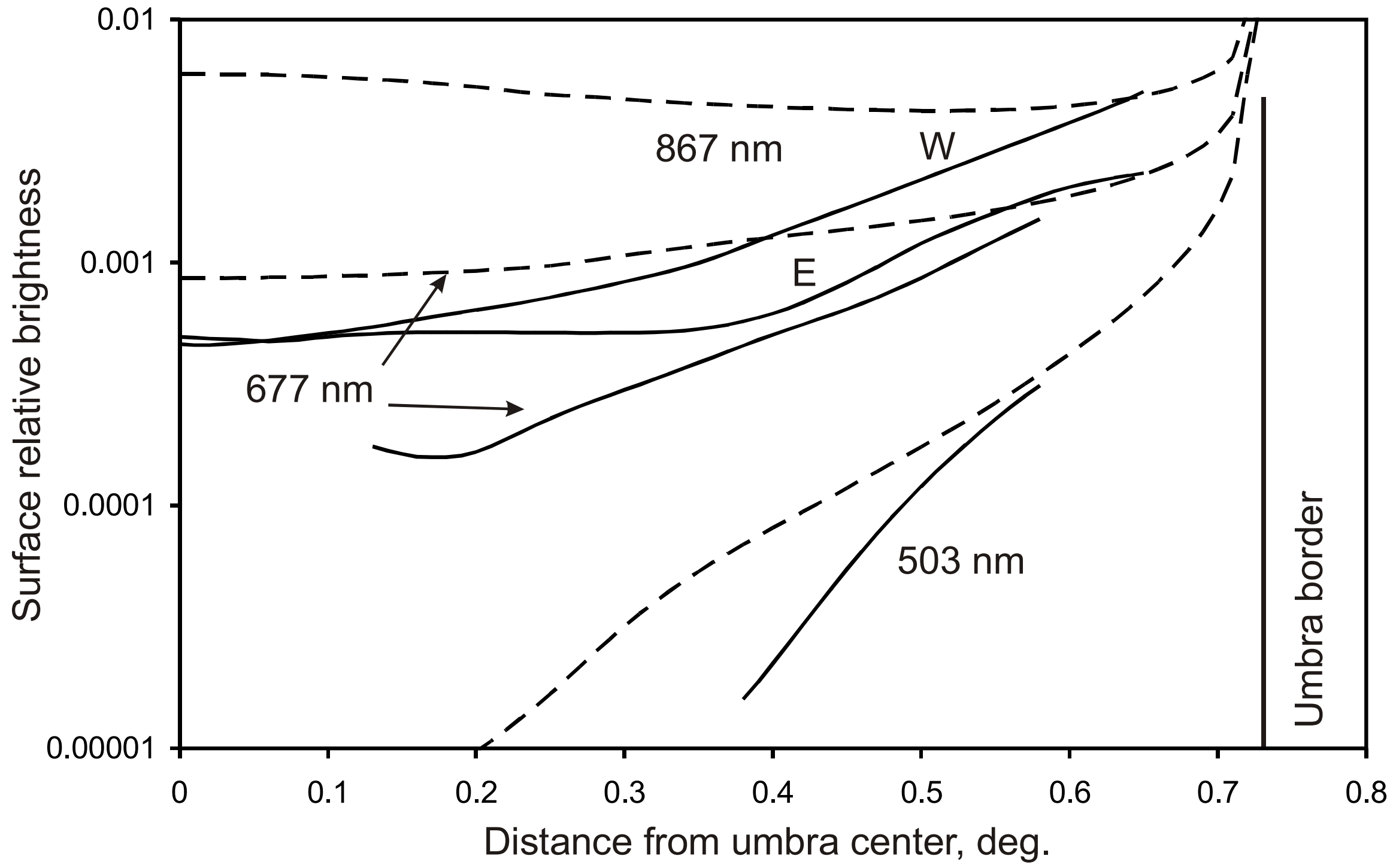

*Figure 2. Theoretical (gaseous atmosphere, dashed lines) and observed brightness of umbra in the directions shown by an arrows in the Figure 1 (W – western direction, E – eastern direction, 30° northwards from the equator). Observational data for 503 nm and 677 nm are in the same western direction.*

The difference between the observational and theoretical brightness values can be used for the calculation of aerosol extinction coefficient at the different altitudes above the Earth's limb. The procedure is described in [4]. The values of aerosol optical depth by the tangent trajectories with different perigee altitudes *h* are calculated. It is the reverse mathematical problem, and the results quality is restricted by the angular size of the Sun [4, 5]. The possible resolution is 0.2° by refraction angle or 3-5 km by the altitude *h*. These optical depth values can be used to find the scattering coefficient of atmospheric aerosol with the same vertical resolution.

Figure 3 shows the dependencies of aerosol extinction coefficient at the altitude 11 km on the limb point coordinates for three wavelengths. This value is defined for both western and eastern limb parts at the wavelength 867 nm, and the troposphere above the eastern (Asian) limb part is obviously more polluted, the same is true for other altitudes.

Western (Atlantic) part of the limb is characterized by the clearer troposphere, but it contains typical equatorial maximum also observed during some previous eclipses [4, 5]. The location of this maximum for June, 2011 eclipse is near the western equatorial African shore. Maximum is seen for all three wavelengths, and the aerosol extinction coefficient at 503 nm is about two times more than at 677 and 867 nm, showing the wavelength dependency of upper troposphere aerosol extinction close to $\lambda^{-1.5}$. Equatorial aerosol maximum is surrounded by two tropical minima with the latitudes equal to ±30°, that is also usual for maritime and western-shore tropical conditions.

Authors would like to thank A. Yudin and A. Tatarnikov for their assistance during the observations preparation and calibration procedures.

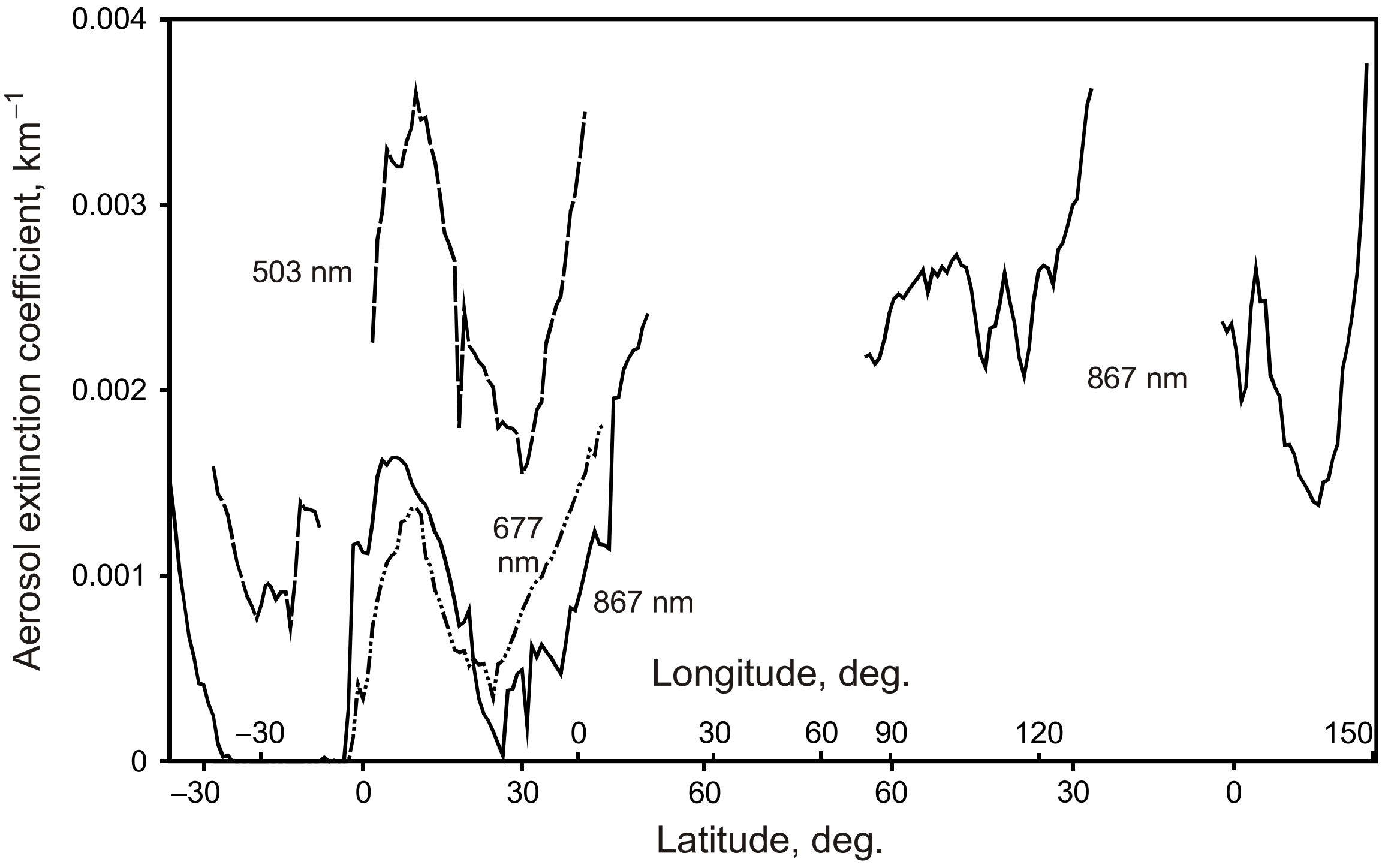

*Figure 3. The values of aerosol extinction coefficients at the altitude 11 km depending on limb point coordinates. The longitude values correspond to the moment 19.25 UT for the western (left) part and the moment 20.46 UT for the eastern part.*